\documentclass[12pt,letterpaper]{article}

\usepackage[T1]{fontenc}
\usepackage[utf8]{inputenc}
\usepackage{lmodern}
\usepackage{microtype}
\usepackage{amsmath,amssymb,mathtools}
\usepackage{booktabs}
\usepackage{array}
\usepackage[margin=1in]{geometry}
\usepackage{graphicx}
\graphicspath{{./}{figures/}}
\usepackage{setspace}
\usepackage[round,authoryear]{natbib}
\usepackage[hidelinks]{hyperref}
\usepackage{url}
\newcommand{\tabnote}[1]{\begin{minipage}{\linewidth}\vspace{4pt}\footnotesize\singlespacing #1\end{minipage}}
\newcommand{\Dref}{D_{\mathrm{ref}}}

\begin{document}

\begin{titlepage}
\begin{center}
\vspace*{0.35cm}
{\Large\bfseries Conditional Deep L\'evy Models for Exotic Derivatives:\\[5pt]
History-Aware Path Generation and P--Q Payoff Diagnostics\par}

\vspace{0.55cm}
{\large Helin Zhao\footnote{Johns Hopkins University. Email:
zhaohelin5@gmail.com. ORCID: 0009-0009-9979-1722. Both authors contributed
equally.}
\qquad
Junchi Shen\footnote{University of Washington, Seattle. Email:
jshen67@uw.edu. ORCID: 0009-0003-9600-9113.}\par}

\vspace{0.25cm}
{October 2026\par}
\end{center}

\vspace{0.35cm}
\begin{abstract}
We develop a history-aware equity-index path generator based on Denoising
L\'evy Probabilistic Models (DLPMs) and evaluate its predictive distributions
through path-dependent derivative payoffs. A conditional U-Net combines
contract state, index identity, 60- and 252-day return histories, and
financial regularization with coordinate-wise stable mixing. A matched
experiment holds initialization, training data, batch order, objective,
computational budget and evaluation checkpoint fixed while comparing
$\alpha=1.9$ with the Gaussian limit $\alpha=2$. On 6{,}824 chronological
windows from eight Chinese and U.S.\ indices, terminal CRPS falls from
0.050585 to 0.048031, a 5.05\% reduction, and path energy improves by 4.03\%.
The paired 60-calendar-day CRPS interval is $[-0.004022,-0.001298]$. Coverage
rises from 56.68\% to 71.00\% while the proper 90\% interval score improves by
20.30\%. Empirical references remain competitive: the full-pool terminal
control attains CRPS 0.047566.

A common P--Q protocol translates the forecasts into vanilla, Asian and
floating-strike lookback outcomes against an independent martingale-corrected
Student-$t$ GARCH benchmark. With quote-error filtering and a fixed-state
initial Delta hedge, the matched Lookback difference is 0.285\% of initial
spot, with a 60-day interval $[0.061\%,0.489\%]$. The evaluation also applies
the same recorded ensembles to accumulator and snowball scenarios. These
contracts expose different responses of early termination and downside
cashflows to the noise specification. The study connects conditional
historical path generation, a controlled stability-parameter comparison,
empirical forecasting references, and payoff-sensitive derivative evaluation.
\end{abstract}

\end{titlepage}

\setcounter{page}{1}
\pagenumbering{arabic}
\doublespacing

\section{Introduction}\label{sec:introduction}
Financial institutions routinely value claims whose cash flows depend not
only on the terminal asset price but also on averages, extrema, observation
schedules, and the order in which market states are visited. Asian and
lookback options are canonical examples; accumulators and autocallable notes
combine path dependence with leverage, early termination, and nonlinear
coupon rules. Their risk cannot be summarized by a point forecast. What
matters is a conditional distribution over entire future paths.

Classical Monte Carlo valuation begins by specifying a stochastic process and
then averaging discounted payoffs. This workflow is transparent, but its
output inherits the restrictions of the assumed process. Geometric Brownian
motion imposes constant volatility and Gaussian increments. Local and
stochastic-volatility models are more flexible, yet they remain parsimonious
parametric representations of a market in which heavy tails, volatility
clustering, drawdowns, trend reversals, and regime-dependent dependence are
empirically prominent \citep{cont2001empirical}. These restrictions are
especially consequential for nonlinear claims: two models can agree on
one-step volatility and still disagree sharply on an Asian average, a running
maximum, or the sequence of states that activates a structured payoff.

Generative models offer a complementary route. Rather than committing to a
small set of diffusion coefficients, one can learn a conditional law of future
returns directly from historical paths. Denoising diffusion models are
attractive because they provide stable likelihood-inspired training, broad
support, and a natural mechanism for conditional scenario generation
\citep{ho2020ddpm,rasul2021timegrad}. A Gaussian forward process, however,
still builds light-tailed noise into the generative mechanism. For financial
returns, this is not an innocuous computational choice. Large moves are rare
but economically decisive, and the reverse process must represent them without
either suppressing tail mass or producing numerically uncontrolled paths.

This paper develops a heavy-tailed alternative based on the Denoising L\'evy
Probabilistic Model of \citet{shariatian2025dlpm}. The DLPM replaces Gaussian
noise with symmetric $\alpha$-stable innovations while retaining a tractable
variance-mixture representation. We combine this mechanism with a
history-aware conditional network and a finance-aware auxiliary objective.
The condition combines contract variables with 60- and 252-day return
sequences, rolling volatility, trend, drawdown, downside risk, reversal
indicators, country, and underlying index. The resulting model is a
conditional physical-measure path generator. It asks: given what was
observable at the start of a contract window, what distribution of future
index paths is plausible?

Nonlinear payoffs make distributional evaluation economically informative.
Small path-distribution differences can change an Asian average or a running
minimum even when vanilla values are similar. We embed the frozen DLPM in a
P--Q quoting experiment. The P side represents the learned real-world belief;
the Q side is an independently fitted Student-$t$ GARCH(1,1) benchmark with a
martingale-corrected drift. Both models value the same payoff. A trade opens
when their valuation gap clears the dealer spread and execution threshold, and
the held-out realized path determines ex-post P\&L. The experiment links
conditional forecasts to realized payoff outcomes.

The controlled comparison is central to interpreting the heavy-tail
mechanism. Two fresh runs share the same initialization, 59{,}240 training
windows, batch sequence, network, condition representation, financial losses
and 16{,}000-update budget. Their fixed EMA-12{,}000 checkpoints are evaluated
on the same 6{,}824 windows with 40 paths each. Changing $\alpha$ changes the
stable-mixture law and its associated scale schedule under common numerical
bounds. This design separates that treatment from changing architecture,
historical information or the regularization recipe. The resulting 5.05\%
CRPS reduction and 4.03\% energy reduction provide controlled evidence within
the chosen financial generator.

Historical references define a second comparison. Against a same-index,
same-tenor nearest-neighbor rule using volatility, trend and drawdown, the
corrected-input original model $\Dref$ improves CRPS by 21.13\% at $K=40$. The
expected score of a random 40-observation historical subset is almost equal to
its score, and the full eligible historical pool scores better. The local
rule, finite-ensemble expectation and full-pool distribution therefore answer
different forecasting questions. Archival Gaussian, stable, parametric,
neural-SDE and foundation-model baselines supply broader methodological
context, while the freshly matched pair supplies the controlled $\alpha$
comparison.

Derivative evaluation probes which distributional differences matter for a
cashflow. Vanilla depends on terminal price, Asian on an arithmetic average,
and Lookback on terminal price relative to the minimum visited. Under the
strict quote-error filter, initial hedging leaves a matched Lookback
difference of 0.285\% of spot, whereas the Vanilla difference interval spans
zero and Asian levels are negative in both arms. Accumulator and snowball
scenarios add repeated observations, leverage and termination rules. Their
cashflows are reported in normalized-notional units alongside event
probabilities, extending the scope beyond the three smooth option contracts.

The contribution is a history-aware financial adaptation of DLPM with a
transparent link from return tokens to conditional price trajectories; a
controlled stability-parameter experiment supported by proper scores,
generated-tail diagnostics and empirical references; and a common evaluation
of terminal, average, extremum and structured-product cashflows.

Section~\ref{sec:literature} reviews the related literature.
Section~\ref{sec:data} defines the data and experiment protocol.
Section~\ref{sec:method} develops Gaussian and stable diffusion and specifies
the deployed implementation. Section~\ref{sec:forecasts} evaluates forecasts;
Section~\ref{sec:payoffs} evaluates option and structured-product payoffs.
Section~\ref{sec:conclusion} discusses the findings and extensions.

\begin{figure}[p]
\centering
\includegraphics[width=\linewidth]{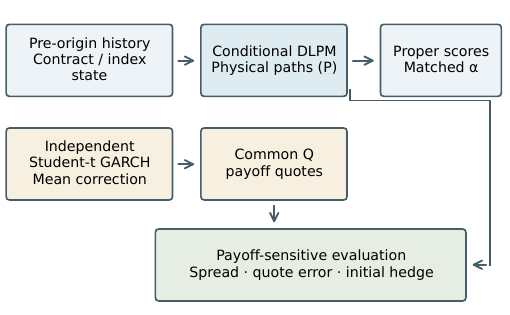}
\caption{Research workflow. Historical conditions generate physical-measure
paths; a separately fitted, martingale-corrected Q benchmark supplies quotes.
Distributional scores and downstream payoffs evaluate the generated paths.}
\label{fig:workflow}
\end{figure}

\section{Related Literature}\label{sec:literature}

\subsection{Pricing structured products and exotic options}

Industry practice relies on four methodologies \citep{deng2014valuation}:
simulation, which averages discounted payoffs over generated price paths and
suits strongly path-dependent products; numerical integration, when payoff
and density admit closed forms; decomposition into bonds, vanilla options,
and simpler exotics; and partial differential equation methods, which capture
continuous state dynamics at high solution cost. Constant-volatility
Black--Scholes dynamics offer a useful starting point, while path-dependent
structures require simulation or an augmented numerical state to track
averages, extrema and contract events. Risk-neutral valuation remains the
common pricing principle across these methods.

A long literature refines these tools: Monte Carlo pricing from
\citet{boyle1977options}, jump-diffusion dynamics from
\citet{merton1976option}, Brownian-bridge quasi-Monte Carlo constructions
\citep{lin2008brownian}, ant colony optimization \citep{kumar2009aco}, and
tensor-network acceleration of lattice pricing \citep{vandamme2025tensor}.
Machine-learning approaches include LSTM option pricing
\citep{pimentel2026lstm}, generative adversarial models for arbitrage-free
implied volatility surfaces \citep{vuletic2025volgan}, and risk-neutral
pricing via GANs \citep{choi2025riskneutral}.

\subsection{Diffusion models for time series}

Denoising diffusion probabilistic models were established by
\citet{ho2020ddpm}, refined by the cosine schedule of
\citet{nichol2021improved}, and accelerated by the deterministic DDIM sampler
of \citet{song2021ddim}. TimeGrad \citep{rasul2021timegrad} pioneered
conditional diffusion for probabilistic time-series forecasting;
DiffSTOCK \citep{daiya2024diffstock} applies relational diffusion to equity
prediction; \citet{meijer2024rise} survey the area.

Financial diffusion research also studies stylized facts in synthetic paths.
\citet{takahashi2025diffusion} use wavelet representations to apply diffusion
to financial time-series images and reconstruct synthetic series. The present
model instead operates directly on scaled return sequences and conditions on
pre-origin historical information. Its financial objective and downstream
payoffs address return dispersion, temporal dependence and path functionals
within that direct conditional representation.

Classical diffusion is Gaussian-driven. For heavy-tailed data,
\citet{shariatian2025dlpm} propose Denoising L\'evy Probabilistic Models
(DLPMs), whose original formulation replaces Gaussian increments with
isotropic $\alpha$-stable noise, $\alpha \in (1,2]$
\citep{samorodnitsky1994stable}. Our financial adaptation uses
\emph{coordinate-wise} stable scale mixing across ordered return coordinates,
instead of one path-level isotropic scale draw. Exploiting the
variance-mixing representation of stable laws, this retains DDPM's
computational structure with tractable conditional Gaussian transitions. The
canonical unbounded $\alpha=2$ innovation is $N(0,2)$; the matched experiment
retains this normalization. Given the well-documented heavy tails of asset
returns \citep{cont2001empirical}, DLPM is a natural generator for financial
paths; we evaluate this financial adaptation through both distributional
forecasting and derivative payoffs.

\subsection{Alternative model families}

Two further families compete in this space. \emph{Neural stochastic
differential equations} parameterize the drift and diffusion of an SDE by
neural networks, trained by likelihood or adversarial criteria
\citep{gierjatowicz2020neural}; imposing no-arbitrage constraints yields
market models suitable for risk-neutral applications
\citep{cohen2023neuralsde}. \emph{Time-series foundation models} such as
Chronos \citep{ansari2024chronos} and TimesFM \citep{das2024timesfm} are
transformers pretrained on billions of observations that produce zero-shot
probabilistic forecasts; their predictive position in a financial application
can be assessed against trained conditional generators and empirical
references. Because both families are natural alternatives to diffusion
generators, our empirical study includes a conditional neural-SDE baseline
recorded on the same historical panel (Section~\ref{sec:forecasts}),
positioning the diffusion results against the classical SDE paradigm rather
than against Gaussian Monte Carlo alone.

\section{Data and Experimental Protocol}\label{sec:data}

\subsection{Markets, dates, and window construction}

The data set covers eight broad equity indices in two large markets:
\begin{center}
CSI~300, CSI~500, CSI~1000, SSE Composite, S\&P~500, NASDAQ, Dow Jones, and
Russell~2000.
\end{center}
The raw collection spans 1 January 2014 through 1 August 2025. Chinese index
levels are obtained from AKShare (tickers 000300, 000905, 000852, and 000001),
and U.S. adjusted index series from Yahoo Finance (\texttt{\^{}GSPC},
\texttt{\^{}IXIC}, \texttt{\^{}DJI}, and \texttt{\^{}RUT}). The series are
price-index or provider-adjusted index levels, not total-return indices.
Chinese rates use SHIBOR 1-, 3-, 6-, and 12-month tenors; U.S. rates use FRED
DGS1MO, DGS3MO, DGS6MO, and DGS1. Rates are matched to contract tenor,
converted to annual decimals, and forward-filled on the local business-day
calendar.

A 252-trading-day pre-start history requirement and complete-target filtering
determine the usable range. Windows have calendar maturities of 30, 90, 180,
and 365 days and record initial price, realized future path, historical
volatility, risk-free rate, country, and index identity.

\begin{table}[ht]
\centering
\caption{Chronological sample construction}
\label{tab:datasplit}
\small\begin{tabular}{lrrrr}
\toprule
Split & Windows & Earliest start & Latest start & Latest target end \\
\midrule
Training & 59{,}240 & 2015-01-02 & 2022-11-11 & 2022-12-13 \\
Validation & 3{,}084 & 2023-01-03 & 2023-10-02 & 2023-11-01 \\
Evaluation & 6{,}824 & 2024-01-02 & 2025-06-06 & 2025-07-06 \\
\bottomrule
\end{tabular}
\tabnote{Counts are after target-end purging. Target end is reconstructed from
the window start, contractual calendar horizon, and stored realized path, and
was checked against the realized-price archive. The evaluation set contains
827 windows for each Chinese index and 879 for each U.S. index.}
\end{table}

The split is chronological and materialized before training. For each
underlying, target end dates are reconstructed from the ordered trading
calendar and stored path length. Target-end purging retains training and
validation windows whose estimated final observation precedes the next split
boundary by at least 20 calendar days. Table~\ref{tab:datasplit} reports the
resulting maximum target dates.

\subsection{Returns, scales, and historical state}

For index level $S_t$, daily log return is
\begin{equation}
r_t=\log(S_t/S_{t-1}),
\end{equation}
and annualized historical volatility over $n$ observations is
\begin{equation}
\sigma_{\mathrm{ann}}
=\sqrt{252}\left(\frac{1}{n-1}\sum_{t=1}^{n}(r_t-\bar r)^2\right)^{1/2}.
\end{equation}
The model represents the target in scaled log-return space,
\begin{equation}
x_{0,t}=r_t/s_{\mathrm{vol}}, \qquad s_{\mathrm{vol}}=0.09,
\end{equation}
and reconstructs levels by exact anchoring at $S_0$:
\begin{equation}
\widehat S_t=S_0\exp\left(s_{\mathrm{vol}}\sum_{j=1}^{t}\widehat x_{0,j}\right).
\label{eq:reconstruct}
\end{equation}
Equation~\eqref{eq:reconstruct} anchors every generated path at the observed
contract level. Variable-tenor windows are padded to length 252 and supplied
with a validity mask; padded entries have zero feature error. The denoising
loss divides by valid-coordinate count, while two auxiliary losses divide by
the fixed tensor size (Section~\ref{sec:deployed}). Missing dated price
observations are dropped. Exchange holidays and suspensions shorten the
recorded path, and payoff evaluation uses its stored
\texttt{actual\_trading\_days} observations.

The numerical condition contains 397 entries. It combines contemporaneous
contract variables with strictly pre-start information: raw 60- and 252-day
return sequences, a 60-day volatility path, 20/60/252-day trends, current and
maximum drawdown, realized and downside volatility, recent reversal,
momentum acceleration, and ratios between short- and long-horizon volatility.
Two categorical identifiers encode country and underlying. Learned embeddings
allow the global model to share common financial structure while retaining
market-specific behavior.

The raw sequences and summary variables play different roles. Sequences
retain local temporal ordering; summaries expose low-frequency state variables
that a convolutional encoder might otherwise learn inefficiently. This
combination is particularly useful at long horizons, where a network must
distinguish a high-volatility rebound from a low-volatility trend even when
their one-day return distributions overlap.

\subsection{Information reconstruction and evaluation scope}

Historical inputs end at the contract origin. For the corrected evaluation,
dated pre-origin prices reconstruct the raw history sequences and their state
features; training-derived outlier bounds and robust scaling are then held
fixed. Independent input checks reproduce conditions after row reordering and
batch-size changes. Perturbing future prices leaves historical conditions
unchanged. Original targets, recorded path lengths, rates, non-history
contract variables and window order are retained.

The original trained weights evaluated with corrected historical conditions
are denoted $\Dref$. The fresh $\alpha=1.9$ and $\alpha=2$ models use these
corrected evaluation conditions and the same frozen training panel. The
evaluation is a retrospective extension of a previously inspected historical
panel, with its windows held outside optimizer training.

\subsection{Matched training and checkpoint protocol}

Both fresh runs use training seed 20260730, batch size 32 and 16{,}000
updates. Adam uses betas $(0.9,0.99)$, with a 2{,}400-update linear
learning-rate warm-up, a 1\% floor, peak rate $10^{-5}$ and cosine decay to
$10^{-7}$. EMA decay is 0.995 and EMA updates occur every ten steps. The
forward schedule has 1{,}000 steps. Initialization, the full training tensor,
batch order and all effective settings are checked between the arms. A
separate CPU shuffle stream seeded with $20261009+e$ at epoch $e$ prevents
different stable-noise draws from changing the examples consumed by training.

The original training recipe evaluated EMA checkpoints at 4{,}000, 8{,}000,
12{,}000 and 16{,}000 steps using validation distributional diagnostics. Its
EMA-12{,}000 checkpoint established the evaluation checkpoint before the fresh
matched runs. Both arms use EMA-12{,}000. Model and checkpoint selection use
validation distributional scores and exclude evaluation and payoff results.

Generation uses 50 skip steps with stochasticity $\eta=1$ and 40 paths per
window. The recorded streams comprise 32 paths with seed 20260730 and eight
with seed 20260803. Auxiliary reverse-mixture chains are cached across batches
of the same shape; the eight-path stream resets its cache at row 5{,}920. The
arms share this recorded cache schedule, conditions and masks.

\subsection{Calendar dependence and uncertainty}

The 6{,}824 windows include 2{,}796 unique asset--date origins. Maturities can
share an origin; adjacent windows share histories and future observations;
indices share calendar shocks. Scores are window-weighted panel means. We
assign origin dates to nonoverlapping 20-, 60-, 90-, 120-, 150-, 180- and
252-calendar-day blocks. A bootstrap replicate samples nonempty block labels
and carries every index and maturity in each selected block together.
Replicate estimates divide resampled value sums by resampled window counts,
preserving the point estimate's weighting. We use 2{,}000 replicates and
percentile 95\% intervals.

Paired comparisons resample window-level differences. Origin-date blocks are
disjoint, but future target paths can overlap across their boundaries. Longer
blocks reduce the number of resampled groups: the panel has 27, 9, 6, 5, 4, 3
and 3 nonempty blocks at the respective resolutions. These intervals summarize
calendar variation within the evaluated historical panel.

\section{Methodology}\label{sec:method}

\subsection{Gaussian conditional diffusion}\label{sec:ddpm}

The forward process of a DDPM is a Markov chain that gradually corrupts the
data $x_0 \in \mathbb{R}^{B \times C \times L}$ with Gaussian noise,
\begin{equation}
q(x_t \mid x_{t-1}) = \mathcal{N}\bigl(x_t;\ \sqrt{1-\beta_t}\,x_{t-1},\
\beta_t I\bigr),
\end{equation}
under the cosine schedule $\beta_t$ of \citet{nichol2021improved} with
$T = 1000$ steps in the final clean-split deployment. With
$\alpha_t = 1 - \beta_t$ and $\bar\alpha_t = \prod_{i \le t} \alpha_i$, the
marginal admits the reparameterization
\begin{equation}
x_t = \sqrt{\bar\alpha_t}\,x_0 + \sqrt{1-\bar\alpha_t}\,\epsilon_t,
\qquad \epsilon_t \sim \mathcal{N}(0, I).
\end{equation}

The reverse process approximates $q(x_{t-1} \mid x_t, x_0)$ by a Gaussian
transition conditioned on contract features $c$,
\begin{equation}
p_\theta(x_{t-1} \mid x_t, c) = \mathcal{N}\bigl(x_{t-1};\
\mu_\theta(x_t, t, c),\ \tilde\beta_t I\bigr),
\end{equation}
where the conditional U-Net (Appendix~\ref{app:architecture}) outputs
$\hat x_0 = f_\theta(x_t, t, c)$ and the posterior mean and variance follow
from Bayes' rule for the Gaussian chain:
\begin{equation}
\tilde\mu_t(x_t, \hat x_0)
= \frac{\sqrt{\bar\alpha_{t-1}}\,\beta_t}{1-\bar\alpha_t}\,\hat x_0
+ \frac{\sqrt{\alpha_t}\,(1-\bar\alpha_{t-1})}{1-\bar\alpha_t}\,x_t,
\qquad
\tilde\beta_t = \frac{1-\bar\alpha_{t-1}}{1-\bar\alpha_t}\,\beta_t .
\end{equation}

The Gaussian family admits the DDIM skip-step sampler
\citep{song2021ddim}: for a subsequence $\{\tau_i\} \subset \{1,\dots,T\}$,
\begin{equation}
x_{\tau_{i-1}}
= \sqrt{\bar\alpha_{\tau_{i-1}}}\,
\underbrace{\frac{x_{\tau_i} - \sqrt{1-\bar\alpha_{\tau_i}}\,
\hat\epsilon_{\tau_i}}{\sqrt{\bar\alpha_{\tau_i}}}}_{\hat x_0}
+ \sqrt{1-\bar\alpha_{\tau_{i-1}} - \sigma_{\tau_i}^2}\;\hat\epsilon_{\tau_i}
+ \sigma_{\tau_i}\, z,
\qquad z \sim \mathcal{N}(0, I),
\end{equation}
with $\sigma_{\tau_i} = 0$ in the fully deterministic case.

\subsection{Training objective I: simple loss}\label{sec:simpleloss}

The \emph{simple} objective is the standard masked denoising regression of
Eq.~\eqref{eq:simple} applied to the network's clean-sequence prediction:
\begin{equation}
\mathcal{L}_{\mathrm{simple}}
= \mathbb{E}_{t,\,x_0,\,\epsilon}
\Bigl[\bigl\| M \odot \bigl(f_\theta(x_t, t, c) - x_0\bigr) \bigr\|_2^2
\big/ \textstyle\sum M \Bigr].
\label{eq:simple}
\end{equation}
This clean-sequence regression describes the simple Gaussian baseline family.
It is distinct from the noise-prediction objective of the deployed matched
DLPM below.

\subsection{Training objective II: composite finance-aware loss}\label{sec:complexloss}

Financial returns exhibit stylized facts---volatility clustering, heavy
tails, skewness, and characteristic spectra---that pure mean-squared error
does not explicitly reward \citep{cont2001empirical}. For the archival
Gaussian finance-aware family, the \emph{composite} objective augments
Eq.~\eqref{eq:simple} with auxiliary regularizers computed between the
predicted clean sequence $\hat x_0$ and the target $x_0$, all masked as in
Eq.~\eqref{eq:simple}:
\begin{align}
L_{\mathrm{jump}} &= \bigl\| (\hat x_{0,i+1} - \hat x_{0,i}) -
(x_{0,i+1} - x_{0,i}) \bigr\|_1
&& \text{(relative jumps)}, \\
L_{\mathrm{vol}} &= \mathrm{SmoothL1}\bigl(\mathrm{Std}(\hat
x_{0,\mathrm{win}}),\ \mathrm{Std}(x_{0,\mathrm{win}})\bigr)
&& \text{(volatility clustering)}, \\
L_{\mathrm{gvol}} &= \bigl|\sigma_{\mathrm{pred}} -
\sigma_{\mathrm{true}}\bigr|
&& \text{(global volatility)}, \\
L_{\mathrm{kurt}} &= \bigl(K_{\mathrm{pred}} - K_{\mathrm{true}}\bigr)^2,
\quad K = \mathbb{E}[z^4] - 3
&& \text{(heavy tails)}, \\
L_{\mathrm{skew}} &= \bigl(\mathbb{E}[z_{\mathrm{pred}}^3] -
\mathbb{E}[z_{\mathrm{true}}^3]\bigr)^2
&& \text{(skewness)}, \\
L_{\mathrm{drift}} &= \Bigl(\textstyle\sum_i \Delta\hat x_{0,i} -
\sum_i \Delta x_{0,i}\Bigr)^2
&& \text{(terminal drift)}, \\
L_{\mathrm{quant}} &= \textstyle\sum_{\tau \in \{0.01, 0.99\}}
L_\tau\bigl(q_\tau(x_0),\ q_\tau(\hat x_0)\bigr)
&& \text{(extreme quantiles)}, \\
L_{\mathrm{spec}} &= \mathrm{SmoothL1}\bigl(\hat P_{\mathrm{norm}},\
P_{\mathrm{norm}}\bigr)
&& \text{(power spectrum)},
\end{align}
where $z$ denotes the standardized sequence, $L_\tau$ is the pinball loss of
\citet{wang2019pinball} at quantile $\tau$, and
$P_{\mathrm{norm}}[k] = P[k]/\max_k P[k]$ is the normalized discrete Fourier
magnitude spectrum. The extreme-quantile pair $\tau \in \{0.01, 0.99\}$
penalizes underestimation of both tails asymmetrically, sharpening the
delineation of downside risk and upside potential.

In this archival Gaussian recipe, each auxiliary term is normalized by an
exponential moving average (EMA) of its own magnitude, and all auxiliary
weights are annealed linearly from zero over a warm-up phase
($s = \min(1, \mathrm{step}/\mathrm{warmup})$):
\begin{equation}
\mathcal{L}_{\mathrm{complex}}
= \mathcal{L}_{\mathrm{simple}}
+ s \sum_{k \in \mathcal{K}} \lambda_k\,
\frac{L_k}{\mathrm{EMA}[L_k] + \varepsilon},
\label{eq:complex}
\end{equation}
with base weights
$(\lambda_{\mathrm{jump}}, \lambda_{\mathrm{vol}}, \lambda_{\mathrm{gvol}},
\lambda_{\mathrm{kurt}}, \lambda_{\mathrm{skew}}, \lambda_{\mathrm{drift}},
\lambda_{\mathrm{quant}}, \lambda_{\mathrm{spec}})
= (0.5, 3.5, 8, 5, 0.5, 2, 3, 2)$.

\subsection{Heavy-tailed diffusion: the DLPM}\label{sec:dlpm}

Gaussian diffusion imposes light-tailed increments on the forward chain. The
original Denoising L\'evy Probabilistic Model of \citet{shariatian2025dlpm} is
isotropic. Our deployed financial-path variant instead uses coordinate-wise
symmetric $\alpha$-stable innovations: for return coordinate $j$,
$\varepsilon^{(\alpha)}_{t,j}\sim\mathcal{S}\alpha\mathcal{S}(1)$,
independently across $j$, with stability index $\alpha\in(1,2]$. This avoids a
single path-level scale draw imposing common shocks across every ordered
return coordinate. The forward chain is applied coordinate-wise,
\begin{equation}
x_{t,j} = \gamma_t x_{t-1,j} + \sigma_t\varepsilon^{(\alpha)}_{t,j},
\qquad t = 1, \dots, T,
\end{equation}
or, equivalently in vector notation,
\begin{equation}
x_t = \gamma_t\, x_{t-1} + \sigma_t\, \varepsilon_t^{(\alpha)},
\qquad t = 1, \dots, T,
\end{equation}
under a \emph{scale-preserving} schedule: with
$\bar\gamma_t = \prod_{i\le t}\gamma_i$, the stability property gives the
marginal
\begin{equation}
x_t = \bar\gamma_t\, x_0 + \bar\sigma_t\, \bar\varepsilon^{(\alpha)},
\qquad
\bar\sigma_t = \bigl(1 - \bar\gamma_t^{\,\alpha}\bigr)^{1/\alpha},
\end{equation}
so $\bar\gamma_t^{\,\alpha} + \bar\sigma_t^{\,\alpha} = 1$ generalizes
variance preservation. We derive $\gamma_t = \alpha_t^{1/\alpha}$ from the
same cosine schedule. The $\alpha$ treatment therefore changes the stable
mixing and the associated scale-preserving coefficients together.

Computations with stable laws become Gaussian conditionally on auxiliary
variables \citep{samorodnitsky1994stable}. In the deployed coordinate-wise
variant, each return coordinate admits the \emph{variance-mixing}
representation
\begin{equation}
\varepsilon^{(\alpha)}_{t,j} \overset{d}{=} \sqrt{a_{t,j}}\; Z_{t,j},
\qquad
a_{t,j} \sim \mathcal{S}_{\alpha/2}^{+},\quad Z_{t,j}\sim\mathcal{N}(0,1),
\label{eq:mixing}
\end{equation}
with the canonical positive mixing law normalized by
\begin{equation}
\mathbb{E}[e^{-ua}]=\exp[-(2u)^{\alpha/2}], \qquad u\ge 0.
\end{equation}
It follows that $\mathbb{E}[e^{iv\sqrt{a}Z}]=e^{-|v|^{\alpha}}$. At
$\alpha=2$, $a=2$ and $\sqrt{a}Z\sim N(0,2)$. The fresh Gaussian-limit arm
retains this normalization and the same objective and bounds. The
$\mathcal{S}_{\alpha/2}^{+}$ distribution is the totally skewed positive
$(\alpha/2)$-stable law. The implementation uses the Chambers--Mallows--Stuck
convention equivalent to stability $\gamma=\alpha/2$, skewness $\beta=1$,
location zero, and scale $c_A=2\cos(\pi\alpha/4)^{2/\alpha}$. For numerical
stability the deployed sampler replaces $a_{t,j}$ by $\min(a_{t,j},100)$.
Consequently, the implemented forward law is a capped stable-mixture
approximation; all reported model and sampler results use this same
convention. Conditionally on the coordinate-wise chain $A=(a_{t,j})_{t,j}$,
the forward process is Gaussian with stochastic diagonal variances. The
following posterior recursion applies element-wise, with $\Sigma_{t,j}$ and
$\Gamma_{t,j}$ replacing the scalar notation:
\begin{equation}
\begin{aligned}
\Sigma_{t,j} &= \sigma_t^2a_{t,j} + \gamma_t^2\Sigma_{t-1,j},\\
\Gamma_{t,j} &= 1 - \gamma_t^2\Sigma_{t-1,j}/\Sigma_{t,j},\\
\tilde m_{t-1,j} &= (x_{t,j} - \bar\sigma_t\Gamma_{t,j}\varepsilon_{t,j})/\gamma_t,\\
\tilde\Sigma_{t-1,j} &= \Gamma_{t,j}\Sigma_{t-1,j}.
\end{aligned}
\label{eq:posterior}
\end{equation}

\paragraph{Theoretical DLPM objective.}
For a marginal sample $x_t=\bar\gamma_t x_0+\bar\sigma_t\sqrt{a}Z$, the
network estimates the aggregate noise
$\varepsilon_t=(x_t-\bar\gamma_t x_0)/\bar\sigma_t$. An ideal stable-noise
model may use an $L_p$ denoising objective
\begin{equation}
\mathcal{L}_p=\mathbb{E}\Bigl[\|M\odot(\varepsilon_\theta(x_t,t,c)-\varepsilon_t)\|_p\Bigr].
\label{eq:dlpmloss}
\end{equation}
For an unbounded stable law with $\alpha<2$, the second noise moment is
infinite. An unsquared Euclidean norm has order-one growth and finite
expectation for $\alpha>1$. For the financial implementation, we use
elementwise Smooth L1 with the reduction rules defined below.

\paragraph{Reverse simulation.}
Conditional on the sampled auxiliary mixture chain, reverse transitions are
diagonal Gaussian transitions with the means and variances in
Eq.~\eqref{eq:posterior}. Skip-step simulation evaluates a subsequence of
noise levels. The deterministic DLIM special case transports a state by
\begin{equation}
x_{\tau_{i-1}}
= \frac{\bar\gamma_{\tau_{i-1}}}{\bar\gamma_{\tau_i}}
\,(x_{\tau_i} - \bar\sigma_{\tau_i}\widehat\varepsilon_{\tau_i})
+ \bar\sigma_{\tau_{i-1}}\widehat\varepsilon_{\tau_i}.
\label{eq:dlim}
\end{equation}
The evaluated 50-step generator uses the stochastic DLIM construction with
$\eta=1$; the deterministic construction corresponds to $\eta=0$.

\subsection{Deployed objective and numerical implementation}\label{sec:deployed}

The backbone is a one-dimensional U-Net \citep{ronneberger2015unet}, with base
width 64, channel multipliers $(1,2,4,8)$, dropout 0.1, and bounded noise
output $2\tanh(\cdot)$. The deployed objective uses masked elementwise Smooth
L1, with transition parameter one:
\begin{equation}
\ell_i=\frac{\sum_j M_{ij}\rho(\varepsilon_{\theta,ij}-\varepsilon_{ij})}
{\sum_j M_{ij}},
\qquad
\rho(u)=
\begin{cases}
u^2/2, & |u|<1,\\[4pt]
|u|-1/2, & |u|\ge 1.
\end{cases}
\end{equation}
The primary term averages $\ell_i$ using regime weights $w_i$ fitted on
training data and capped at 1.5. These weights balance volatility, trend, and
drawdown states within indices. Smooth L1 gives linear growth at large errors,
suited to the heavy-tailed denoising target.

An auxiliary loss acts on
$\widehat x_0=(x_t-\bar\sigma_t\varepsilon_\theta)/\bar\gamma_t$. It combines
pointwise and cumulative returns; terminal drift; volatility and
absolute-return magnitude; serial correlation, volatility clustering, and
leverage; tail quantiles and exceedances; drawdown; and cumulative-path
quantiles. Fixed component weights are shared between arms. The complete
objective is
\begin{equation}
\mathcal{L}=\frac{\sum_i w_i\ell_i}{\sum_i w_i}
+0.07\min(s/1000,1)\sum_m\lambda_m\mathcal{R}_m,
\label{eq:deployedloss}
\end{equation}
where $s$ is the optimizer step. Table~\ref{tab:regularizers} gives the fixed
component coefficients shared by both arms. Reconstruction and
cumulative-path terms constrain the transition from return tokens to prices;
dispersion, dependence, and tail terms constrain the distributional features
relevant to nonlinear payoffs. These components augment the denoising
objective under its 1{,}000-update auxiliary warm-up.

\begin{table}[ht]
\centering
\caption{Fixed financial regularizer coefficients $\lambda_m$ in
Eq.~\ref{eq:deployedloss}. Entries separated by slashes follow the feature
order in each row.}
\label{tab:regularizers}
\small\begin{tabular}{lr}
\toprule
Feature discrepancies & Coefficients \\
\midrule
Return reconstruction / range & 2.5 / 5 \\
Terminal / cumulative tokens & 0.7 / 0.7 \\
Volatility / mean absolute token & 3 / 5 \\
Rolling volatility / absolute-return ACF & 0.5 / 0.35 \\
Signed ACF: lag 1 / lag 5; leverage & 0.2 / 0.1 / 0.1 \\
Signed / absolute tails; tail-to-IQR & 0.1 / 4.5 / 0.05 \\
Soft exceedances: 0.5/1/2 & 1 / 0.6 / 0.2 \\
Maximum drawdown & 0.35 \\
Signed / absolute / terminal log-path quantiles & 0.5 / 0.75 / 0.9 \\
Terminal log-return per observation & 0.35 \\
\bottomrule
\end{tabular}
\tabnote{Both fresh arms use all 22 coefficients, the outer scale 0.07, and
the 1{,}000-update auxiliary warm-up. Token units are scaled log returns;
log-path units multiply their cumulative sum by 0.09.}
\end{table}

For the auxiliary terms, let
$z_j=M_j\operatorname{clip}(\widehat x_{0,j},-2.5,2.5)$ and apply the same
transformation to the training target. For these terms, $M_0=0$ sets the
initial return token to zero. We distinguish cumulative scaled tokens
$C_j=\sum_{\ell\le j}z_\ell$ from cumulative log returns $L_j=0.09C_j$.
Feature discrepancies use Smooth L1, with a squared range penalty and soft
exceedance rates
\begin{equation}
\begin{aligned}
\mathcal{R}_{\mathrm{range}}&=\frac{\sum_j M_j(|z_j|-1.25)_+^2}{\sum_j M_j},\\
e_c&=\operatorname{mean}_{M_j=1}\operatorname{sigmoid}(8(|z_j|-c)),
\end{aligned}
\end{equation}
where $c\in\{0.5,1,2\}$. Range sums include the batch, while exceedances and
token quantiles pool valid batch coordinates. Signed tail quantiles use
probabilities $(.01,.99)$; absolute-token quantiles use
$(.50,.75,.90,.95,.99)$. The tail-to-IQR ratio is
$(q_{.95}-q_{.05})/(q_{.75}-q_{.25})$, with denominator floor $10^{-6}$.

Within each target, volatility uses the population standard deviation,
rolling volatility uses 32-token windows with stride 16, and dependence terms
compare signed-return correlations at lags one and five, lag-one
absolute-return correlation, and correlation of a return with the next
absolute return. Variance and correlation calculations use $10^{-6}$
stabilization. Rolling-volatility errors average all batch subwindows, with
empty subwindows contributing zero. Maximum drawdown is computed from
$p_j=\exp(\operatorname{clip}(L_j,-5,5))$ as
$\min_{j:M_j=1}\{p_j/\max_{0\le\ell\le j}p_\ell-1\}$. The running maximum
includes $p_0=1$; the minimum excludes the anchor and padding. Signed log-path
and terminal log-return quantiles use $(.01,.05,.50,.95,.99)$; absolute
log-path quantiles use $(.50,.90,.95,.99)$. Path quantiles pool valid
coordinates, terminal quantiles pool window endpoints, and the slope term
compares $L_n/n$. Pointwise and cumulative-token Smooth L1 losses sum
valid-coordinate errors and divide by the total number of batch-tensor
elements, including the zero-error anchor and padding. These training-path
penalties specify the generator before downstream contract evaluation.

\paragraph{Bounded sampling.}
Numerical implementation is part of the model specification. Training mixture
variables and the sampled reverse mixture chain are capped at $a\le 100$.
Training targets $\sqrt{a}Z$ are not individually clipped at two; the
prediction network output is $2\tanh(\cdot)$, in $[-2,2]$. During generation,
initial noise from the stable generator is clipped to $[-2,2]$. Denoised
return tokens and final output are bounded by $|\widehat x_0|\le 1.35$, and
intermediate states have a time-dependent trust bound
$8\bar\gamma_t+2\bar\sigma_t$. Financial-loss calculations additionally use a
2.5-unit input bound and a 1.25-unit return-range penalty threshold.

Generation uses the DLIM skip-step construction with stochasticity $\eta=1$;
deterministic DLIM is the $\eta=0$ special case. The evaluated process is a
bounded stable-mixture generator. Section~\ref{sec:forecasts} assesses its
generated tails and financial consequences. All bounds and the sampler
schedule are identical across the matched arms. The $\alpha$ treatment
changes the mixture distribution and its associated scale-preserving
coefficients under this fixed numerical protocol.

\section{Model Comparison and Distributional Evaluation}\label{sec:forecasts}

\subsection{Comparison design and proper scores}

The fresh pair compares stable and Gaussian-limit noise within one
architecture. $\Dref$ supplies the original-model reference under corrected
conditions. The archival landscape uses recorded training/input protocols,
including three training seeds for neural control families. Those seed means
characterize the recorded families; they are separate from the one-seed
matched experiment. Historical terminal pools supply empirical controls at
several ensemble sizes.

For a terminal normalized price $Y=S_n/S_0$ and ensemble
$X_1,\ldots,X_M$, empirical CRPS is
\begin{equation}
\mathrm{CRPS}=\frac{1}{M}\sum_m|X_m-Y|-\frac{1}{2M^2}\sum_{m,m'}|X_m-X_{m'}|.
\end{equation}
It rewards concentration around the realized value while accounting for
ensemble dispersion \citep{gneiting2007proper}. Fair CRPS replaces $M^2$ by
$M(M-1)$ in the pair-distance denominator. We retain both definitions in the
archival table to expose finite-ensemble effects. The fresh arms have the
same $M=40$ and are compared using empirical CRPS.

For complete normalized trajectories $\mathbf{X}_m$ and realization
$\mathbf{Y}$, the path energy score is
\begin{equation}
\mathrm{Energy}=\frac{1}{M}\sum_m\|\mathbf{X}_m-\mathbf{Y}\|_2
-\frac{1}{2M^2}\sum_{m,m'}\|\mathbf{X}_m-\mathbf{X}_{m'}\|_2.
\end{equation}
Distances include the recorded path observations. Energy is not divided by
the square root of tenor; comparisons use the same windows and are also
reported by maturity. Central-90\% terminal coverage uses ensemble quantiles
$l=q_{.05}$ and $u=q_{.95}$. The corresponding proper interval score is
\begin{equation}
\mathrm{IS}_{.1}(l,u;Y)=u-l+20(l-Y)_+ +20(Y-u)_+.
\end{equation}
Coverage and width describe the forecast interval separately; the interval
score assesses their trade-off. Pointwise fan coverage averages daily
interval inclusion within each window, then averages windows. It is distinct
from a simultaneous whole-path coverage probability. The volatility ratio
averages generated within-path return standard deviations relative to the
realized within-window standard deviation.

\subsection{Matched stable versus Gaussian-limit forecasts}

The $\alpha=1.9$ arm attains CRPS 0.048031 and energy 0.334585, compared with
0.050585 and 0.348646 for $\alpha=2$. The relative reductions are 5.05\% and
4.03\%. Wider intervals raise terminal coverage by 14.32 percentage points,
from 56.68\% to 71.00\%, while the interval score improves by 20.30\%. The
joint improvement connects added predictive dispersion with fewer interval
misses.

\begin{table}[ht]
\centering
\caption{Common-panel path quality: 6{,}824 windows and 40 paths per window.
The fresh arms are the matched pair; $\Dref$ is the earlier model with
corrected inputs.}
\label{tab:pathquality}
\small\begin{tabular}{lrrr}
\toprule
Metric & $\alpha=1.9$ & $\alpha=2$ & $\Dref$ \\
\midrule
CRPS & 0.04803 & 0.05059 & 0.04872 \\
Path energy & 0.33458 & 0.34865 & 0.33942 \\
90\% coverage & 71.00\% & 56.68\% & 70.34\% \\
90\% width & 0.16706 & 0.11493 & 0.16338 \\
90\% interval score & 0.48761 & 0.61181 & 0.48684 \\
Volatility ratio & 0.88561 & 0.60318 & 0.88088 \\
\bottomrule
\end{tabular}
\tabnote{CRPS, energy, and interval score: lower is better. All prices are
normalized by initial spot. Coverage is terminal; volatility ratio is
generated/realized. Nominal coverage is 90\%.}
\end{table}

\begin{table}[ht]
\centering
\caption{Matched path-score differences across calendar-block lengths}
\label{tab:matchedblocks}
\small\begin{tabular}{rrrr}
\toprule
Block days & Blocks & CRPS difference CI & Energy difference CI \\
\midrule
20 & 27 & [-0.00353, -0.00170] & [-0.01832, -0.00998] \\
60 & 9 & [-0.00402, -0.00130] & [-0.01898, -0.00790] \\
90 & 6 & [-0.00385, -0.00134] & [-0.02014, -0.00758] \\
120 & 5 & [-0.00412, -0.00118] & [-0.01937, -0.00674] \\
150 & 4 & [-0.00384, -0.00097] & [-0.01958, -0.00787] \\
180 & 3 & [-0.00417, -0.00112] & [-0.01983, -0.00974] \\
252 & 3 & [-0.00382, 0.00130] & [-0.01513, 0.00449] \\
\bottomrule
\end{tabular}
\tabnote{All intervals are percentile 95\% intervals for $\alpha=1.9$ minus
$\alpha=2$, from 2{,}000 common-calendar replicates. Point differences are
$-0.002554$ and $-0.014061$, respectively. Negative differences favor the
stable arm.}
\end{table}

The paired CRPS difference is $-0.002554$, with 60-day interval
$[-0.004022,-0.001298]$. The energy difference is $-0.014061$, with interval
$[-0.018983,-0.007897]$. CRPS intervals lie below zero for blocks from 20
through 180 days. The 252-day interval, based on three nonempty blocks, spans
zero.

Both countries show greater stable-arm coverage, with 60.52\% in China and
80.86\% in the United States. The group breakdowns also expose departures from
the aggregate ordering. At the 365-day tenor, the 116 windows yield CRPS
0.055155 for $\alpha=1.9$ versus 0.048214 for $\alpha=2$. Dow Jones is almost
tied, with a slightly lower Gaussian-arm point estimate. The shorter
maturities dominate the window-weighted panel. Conditional forecast
comparisons therefore depend on the market and maturity mix as well as the
noise mechanism.

\begin{figure}[ht]
\centering
\includegraphics[width=0.89\linewidth]{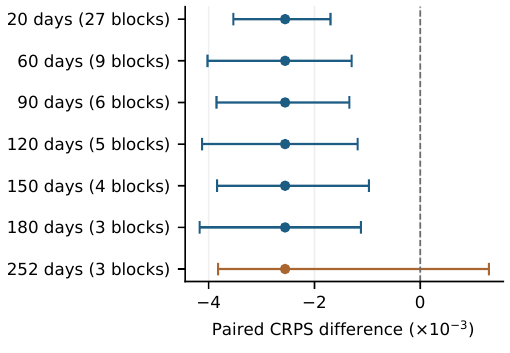}
\caption{Matched terminal CRPS differences across common calendar
resolutions. Negative values favor $\alpha=1.9$; bars show 95\% intervals and
labels give the number of nonempty blocks.}
\label{fig:matchedblocks}
\end{figure}

\begin{table}[ht]
\centering
\caption{Matched terminal forecasts by market, index, and maturity}
\label{tab:groups}
\small\begin{tabular}{lrrrrr}
\toprule
Group & $N$ & CRPS 1.9 & CRPS 2 & Coverage 1.9 & Coverage 2 \\
\midrule
China & 3308 & 0.06169 & 0.06602 & 60.52 & 44.77 \\
United States & 3516 & 0.03518 & 0.03606 & 80.86 & 67.89 \\
CSI 1000 & 827 & 0.08662 & 0.09226 & 48.00 & 35.91 \\
CSI 300 & 827 & 0.04738 & 0.05108 & 68.80 & 48.25 \\
CSI 500 & 827 & 0.07158 & 0.07678 & 51.39 & 37.97 \\
Dow Jones & 879 & 0.02678 & 0.02675 & 88.62 & 77.13 \\
NASDAQ & 879 & 0.04361 & 0.04467 & 73.15 & 58.59 \\
Russell 2000 & 879 & 0.04058 & 0.04289 & 74.40 & 61.32 \\
S\&P 500 & 879 & 0.02976 & 0.02993 & 87.26 & 74.52 \\
SSE Composite & 827 & 0.04117 & 0.04398 & 73.88 & 56.95 \\
30 days & 2796 & 0.03359 & 0.03528 & 68.99 & 54.69 \\
90 days & 2316 & 0.05707 & 0.06082 & 67.70 & 52.37 \\
180 days & 1596 & 0.05970 & 0.06271 & 77.69 & 64.35 \\
365 days & 116 & 0.05515 & 0.04821 & 93.10 & 85.34 \\
\bottomrule
\end{tabular}
\tabnote{Coverage entries are percentages for the central 90\% terminal
interval. All groups use common windows, 40 paths and unchanged trained
checkpoints. Group scores are descriptive summaries, with no subgroup-based
parameter selection.}
\end{table}

\clearpage
\begin{table}[t]
\singlespacing
\centering
\caption{First-return extremes under common realized thresholds}
\label{tab:tails}
\small\begin{tabular}{lrrrrr}
\toprule
Distribution & Left (\%) & Right (\%) & Left/IQR & Right/IQR & Excess kurtosis \\
\midrule
Realized & 1.011 & 1.011 & 2.932 & 3.333 & 9.750 \\
$\Dref$ & 0.621 & 0.651 & 2.920 & 2.613 & 14.963 \\
$\alpha=1.9$ & 0.661 & 0.651 & 2.884 & 2.550 & 13.810 \\
$\alpha=2$ & 0.064 & 0.028 & 2.175 & 1.863 & 1.652 \\
\bottomrule
\end{tabular}
\tabnote{Left/right exceedances use the realized 1\% and 99\% cutoffs.
Tail-to-IQR uses the distance from the median to the 1\% or 99\% quantile,
divided by the interquartile range. Origins repeat across maturities, so these
are window-weighted descriptive summaries of bounded outputs.}
\end{table}

\subsection{Generated tails and implementation sensitivity}

First-return extremes use fixed realized 1\% and 99\% cutoffs of $-3.619\%$
and $4.217\%$. Stable-arm exceedances are 0.661\% on the left and 0.651\% on
the right, compared with 0.064\% and 0.028\% for the Gaussian arm. Its
tail-to-IQR ratios are also closer to the realized first-return panel. The
heavy-tail mechanism thus remains visible in the empirically bounded outputs.
The finite-support generator's diagnostic tails should be interpreted at
these observed thresholds, rather than identified with an asymptotic stable
exponent.

\begin{table}[ht]
\singlespacing
\centering
\caption{Fixed-subset sampler and context diagnostics}
\label{tab:sensitivity}
\small\begin{tabular}{lrrrr}
\toprule
Variant & CRPS & Coverage (\%) & Width & Interval score \\
\midrule
$\Dref$, baseline & 0.04152 & 78.71 & 0.1658 & 0.3589 \\
Initial-noise bound 4 & 0.04152 & 78.71 & 0.1658 & 0.3589 \\
Token bound 2 & 0.04142 & 80.08 & 0.1773 & 0.3527 \\
Joint relaxation & 0.04137 & 81.84 & 0.1862 & 0.3462 \\
Typical donor history & 0.04475 & 75.59 & 0.1674 & 0.3906 \\
Volatile donor history & 0.04556 & 76.37 & 0.1705 & 0.4277 \\
\bottomrule
\end{tabular}
\tabnote{The same 512 prespecified windows and original trained model are
used throughout. Joint relaxation changes initial-noise, mixture and
denoised-token bounds; donor-history rows substitute within-index, same-tenor
training contexts. The fresh matched arms retain the original fixed bounds.}
\end{table}

The sensitivity subset contains 512 windows: four 16-row batches per index at
fixed positions chosen without score filtering. With the original trained
model, increasing the initial-noise clip alone leaves coverage at 78.71\%.
Increasing the denoised-token bound raises it to 80.08\%; jointly relaxing the
initial-noise bound to four, mixture cap to 200 and token bound to two raises
coverage to 81.84\% and lowers interval score from 0.3589 to 0.3462. The
paired CRPS interval spans zero, and longer calendar blocks weaken the
coverage-gain interval evidence. These diagnostics characterize the role of
numerical bounds; the matched models retain the fixed production bounds.

Typical and volatile same-index, same-tenor training histories provide two
context substitutions while contract state and index identity remain fixed.
Both worsen CRPS relative to the subset's supplied-history baseline. They
measure the trained network's response to historical input, a different
question from retraining an architecture without history.

\subsection{Historical controls and neighborhood size}

For each window, the local control selects same-index, same-tenor training
terminal returns using squared distance in training-standardized 60-day
volatility, trend and current drawdown. We evaluate
$K=10,20,40,80,160,320,640$ and the full eligible pool. The reported grid
measures sensitivity to the local rule's neighborhood size; the reference
comparison remains at $K=40$.

Let an eligible pool contain $N$ terminal values $Z_1,\ldots,Z_N$. For a
uniform subset of $K$ distinct values, the exact expected empirical CRPS is
\begin{equation}
\mathbb{E}[\mathrm{CRPS}_K]=\frac{1}{N}\sum_{a=1}^{N}|Z_a-Y|
-\frac{K-1}{2K}\,\frac{\sum_{a\ne b}|Z_a-Z_b|}{N(N-1)}.
\end{equation}
The expression follows from the equal marginal inclusion probabilities of a
single value and an ordered pair. It separates neighborhood selection from
finite-ensemble size without adding Monte Carlo sampling error.
\begin{table}[t]
\centering
\caption{Neighborhood sensitivity of the empirical terminal-return controls.
Gain uses $\Dref$ CRPS 0.048725 against the state control.}
\label{tab:neighborhood}
\small\begin{tabular}{rrrr}
\toprule
$K$ & State CRPS & Pooled CRPS & Gain \\
\midrule
10 & 0.07225 & 0.05264 & 32.56\% \\
20 & 0.06640 & 0.05009 & 26.62\% \\
40 & 0.06178 & 0.04881 & 21.13\% \\
80 & 0.05749 & 0.04818 & 15.25\% \\
160 & 0.05436 & 0.04786 & 10.36\% \\
320 & 0.05199 & 0.04770 & 6.29\% \\
640 & 0.04934 & 0.04762 & 1.24\% \\
all & 0.04757 & 0.04757 & -2.44\% \\
\bottomrule
\end{tabular}
\tabnote{Pooled is the exact expected same-$K$ uniform-subset CRPS.
\emph{all} uses the full same-index, same-tenor training pool. This reference
differs from the archival path-block bootstrap.}
\end{table}
At $K=40$,
$\Dref$ has CRPS 0.048725 versus 0.061777 for the local control, an
improvement of 21.13\%. Its score is almost equal to the same-size pooled
expectation, 0.048815. The local-control advantage shrinks to 6.29\% at
$K=320$, where the 60-day paired interval spans zero. The full-pool terminal
distribution scores 0.047566, 2.44\% below $\Dref$ and also below the fresh
stable arm. This locates the generator's advantage relative to a specific
local selection rule rather than all empirical forecasting distributions.

\subsection{Broader archival benchmark landscape}

The archival comparator families include contract-only stable diffusion,
Gaussian diffusion under simple and finance-aware objectives, parametric GBM,
learned continuous-time dynamics, general time-series forecasting and complete
historical-path resampling. Their recorded objectives and training/input
protocols are retained; the fresh pair supplies the within-recipe $\alpha$
contrast. The table reports terminal 90\% coverage.

Historical path bootstrap attains energy 0.329656 and terminal coverage
88.80\%. Together with the full-pool terminal control, it establishes
historical resampling as a competitive reference. The bootstrap archive has
32 paths, while the fresh arms have 40. The fresh pair supplies the matched
treatment comparison.

\begin{table}[ht]
\centering
\caption{Archival comparator landscape rescored on 6{,}824 windows using
common metric definitions. $M$ is the path count. Coverage is terminal 90\%
coverage; the three loss scores are lower-is-better.}
\label{tab:archival}
\small\begin{tabular}{lrrrrr}
\toprule
Recorded model & $M$ & CRPS & Fair CRPS & Energy & Coverage (\%) \\
\midrule
History-aware DLPM (archive) & 40 & 0.04851 & 0.04775 & 0.33680 & 70.43 \\
Contract-only DLPM & 32 & 0.05483 & 0.05370 & 0.36261 & 72.42 \\
Gaussian DDPM, simple & 32 & 0.08307 & 0.08245 & 0.54608 & 27.05 \\
Gaussian DDPM, finance-aware & 32 & 0.15207 & 0.15139 & 0.91529 & 21.64 \\
Historical path bootstrap & 32 & 0.04861 & 0.04687 & 0.32966 & 88.80 \\
Historical-drift GBM & 32 & 0.04983 & 0.04800 & 0.33968 & 91.91 \\
Conditional neural SDE & 32 & 0.05786 & 0.05652 & 0.37742 & 71.79 \\
Chronos-T5, zero-shot & 32 & 0.05418 & 0.05296 & 0.36675 & 74.06 \\
\bottomrule
\end{tabular}
\tabnote{Neural control families average three recorded training seeds;
archival history-aware DLPM is one frozen model. These recorded-input archives
use their original training and ensemble protocols. Corrected-input $\Dref$
and fresh matched arms are reported separately in Table~\ref{tab:pathquality}.
Fair CRPS uses $M(M-1)$ in place of $M^2$ in the ensemble-distance term.}
\end{table}

\subsection{Original-model provenance and conditional fans}

The original recorded-input archive had CRPS 0.048512 and energy 0.336799.
With corrected historical conditions and training-fixed preprocessing,
$\Dref$ has CRPS 0.048725 and energy 0.339420. The associated $K=40$
local-control comparison changes from an archival 23.35\% reduction to
21.13\%. These input audits use the same original trained weights and recorded
streams. They are distinct from the fresh 5.05\% alpha-effect estimate.

The fans illustrate cross-index variation in conditional paths. Calibration,
tail frequencies and subgroup differences are measured over the complete
numerical panels above and linked to the nonlinear contracts evaluated next.

\begin{figure}[p]
\centering
\includegraphics[height=0.88\textheight,keepaspectratio]{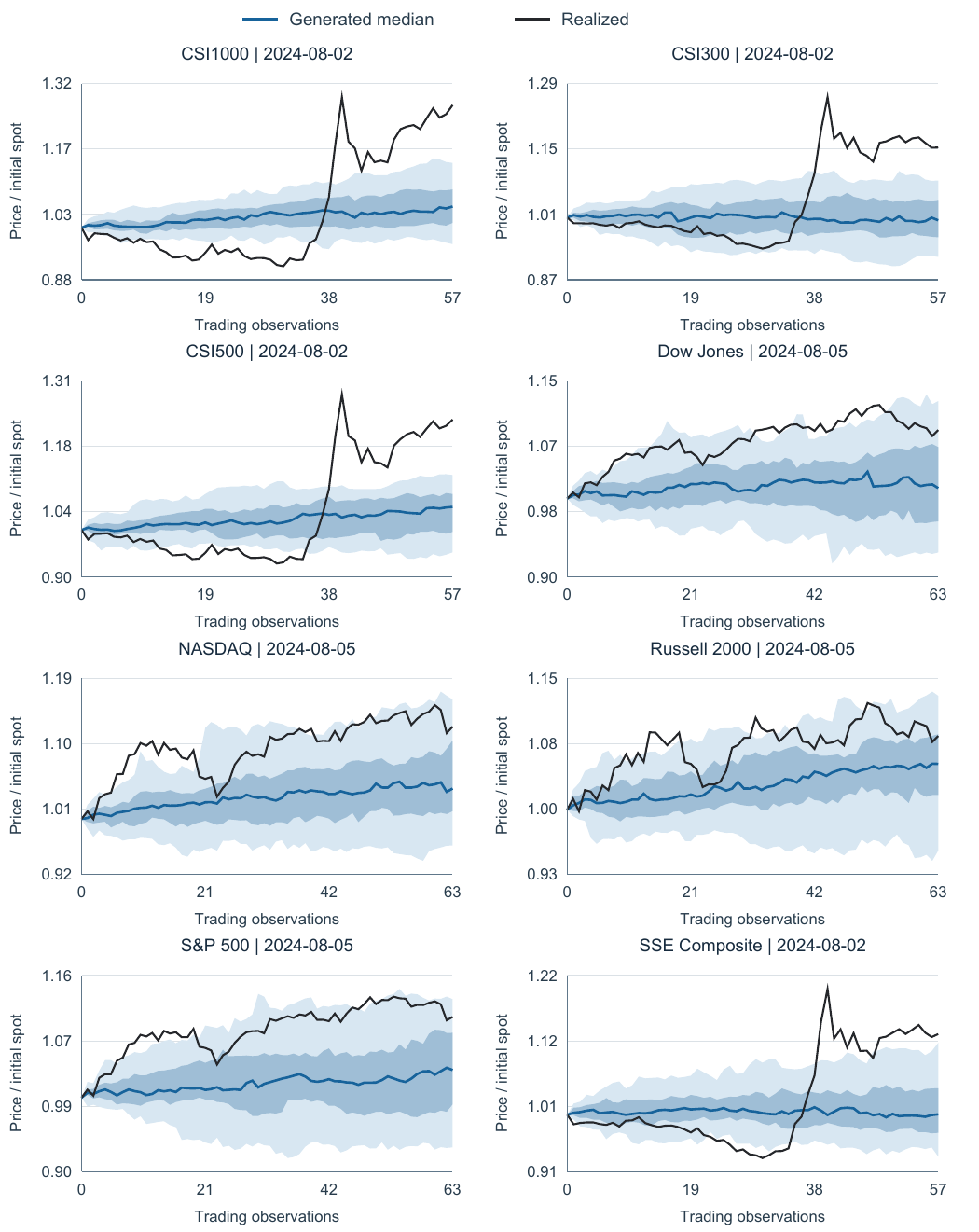}
\singlespacing
\caption{Corrected-input original-model conditional fans for eight indices.
Each panel uses the middle dated 90-day window for its index. Generated and
realized levels share initial spot.}
\label{fig:fans}
\end{figure}

\section{Path-Dependent Derivatives and the P--Q Evaluation}\label{sec:payoffs}

\subsection{An independently constructed risk-neutral benchmark}

For each index, the Q benchmark begins with a Student-$t$ GARCH(1,1)
volatility recursion \citep{bollerslev1986garch},
\begin{align}
h_{t+1} &= \omega+\alpha \epsilon_t^2+\beta h_t,\\
z_{t+1} &\sim t_\nu,\qquad \epsilon_{t+1}=\sqrt{h_{t+1}}z_{t+1}.
\end{align}
Parameters are fitted to a deduplicated chronological return series through
the validation period. The GARCH recursion represents volatility persistence,
while Student-$t$ innovations allow heavy-tailed shocks.

To construct Q from the fitted variance process, we specify its innovation
law and martingale-corrected conditional drift. Let $z$ denote a standardized
Student-$t$ draw. A symmetric raw cap is solved so that, after truncation and
re-standardization,
\begin{equation}
\widetilde z
=\frac{\operatorname{clip}(z,-c_\nu,c_\nu)}
{\sqrt{\operatorname{Var}[\operatorname{clip}(z,-c_\nu,c_\nu)]}},
\qquad |\widetilde z|\leq 8,\qquad
\operatorname{Var}(\widetilde z)=1.
\end{equation}
Post-truncation re-standardization preserves unit innovation variance and the
fitted GARCH conditional variance.

The GARCH model is fitted to daily log returns (temporarily expressed in
percentage points for numerical estimation and converted back afterward).
Accordingly, $h_{t+1}$ is a one-trading-day conditional log-return variance,
not an annualized variance. For annualized risk-free rate $r_t$, annualized
dividend or carry yield $q_t$, and $\Delta=1/252$, define
\begin{equation}
\kappa_t=\log\mathbb E\left[
\exp\left(\sqrt{h_{t+1}}\,\widetilde z_{t+1}\right)
\mid\mathcal F_t\right].
\end{equation}
The RN-Q level process is
\begin{equation}
S_{t+1}=S_t\exp\left((r_t-q_t)\Delta-\kappa_t+
\sqrt{h_{t+1}}\,\widetilde z_{t+1}\right).
\label{eq:rnq}
\end{equation}
By construction,
\begin{equation}
\mathbb E^Q[e^{-(r_t-q_t)\Delta}S_{t+1}\mid\mathcal F_t]=S_t.
\end{equation}
The log-mgf is evaluated numerically under the same capped, unit-variance
innovation law used for simulation. Empirical audits across all indices and
20-, 60-, 120-, and 240-day horizons produce mean discounted-price ratios close
to one. Independent quadrature checks place the daily conditional-mean
correction error below $2.2\times10^{-6}$ for volatility up to 5\%; mean
discounted terminal ratios by index range from 0.999647 to 1.000210.

The benchmark sets $q_t=0$ for the price-index underlyings. Q is fitted to
historical returns through the validation period and is a model-based,
zero-carry risk-neutral benchmark; it is not calibrated to market option
quotes. The numerical martingale checks test this specified innovation law.

\subsection{Payoff functionals and the common quoting protocol}

P and Q value the same recorded contract state and observation dates. For a
window with $n$ future trading observations, the discount factor is
$D=\exp(-rn/252)$, and the contracts are
\begin{equation}
\begin{aligned}
H_{\mathrm V}(S_{0:n})&=(S_n-S_0)_+,\\
H_{\mathrm A}(S_{0:n})&=\left(\frac{1}{n+1}\sum_{j=0}^{n}S_j-S_0\right)_+,\\
H_{\mathrm L}(S_{0:n})&=S_n-\min_{0\le j\le n}S_j.
\end{aligned}
\end{equation}
Vanilla tests terminal price, Asian the arithmetic average, and Lookback the
terminal price relative to the running minimum. The average and minimum
include the initial observation. P has 40 paths per window and Q has 256. The
estimated values $V_P$ and $V_Q$ are Monte Carlo means of $DH$. All models
share Q paths, rates, realized cashflows, observation dates and initial hedge
states. Unique asset--window joins are checked against spot and recorded path
length.

With total spread $g=0.05$, the dealer ask and bid are $A=1.025V_Q$ and
$B=0.975V_Q$. A materiality and quote-error hurdle is
\begin{equation}
h_k=0.01S_0+k\sqrt{\mathrm{SE}_P^2+\mathrm{SE}_Q^2},\qquad k\in\{0,1.96\}.
\end{equation}
The P side buys if $V_P-A>h_k$, sells if $B-V_P>h_k$, and otherwise records no
trade. Standard errors are nominal within-ensemble discounted-payoff standard
errors. The stricter rule assesses whether the estimated valuation
disagreement clears quote noise in addition to the fixed materiality
threshold. The same rule applies to every contract and arm.

Raw normalized P\&L is $(DH^{\mathrm{realized}}-A)/S_0$ for a buy and
$(B-DH^{\mathrm{realized}})/S_0$ for a sell. Every no-trade window contributes
zero. Means include all 6{,}824 windows; execution rates and payoff means
therefore remain separate outcomes. This accounting includes the option
bid--ask spread and, for the hedge below, financing at the window rate.

\subsection{Fixed-state initial Delta and payoff accounting}

The initial hedge is established immediately after the observed origin
fixing. Let $z_0$ denote the observed contract state. Common-path central
differences with relative bump $b=10^{-3}$ estimate
\begin{equation}
\Delta_Q=\frac{V_Q(S_0(1+b);z_0)-V_Q(S_0(1-b);z_0)}{2bS_0}.
\end{equation}
Strike, the initial Asian fixing and the initial Lookback minimum stay fixed;
only future simulated observations are rescaled. Holding the observed state
fixed is necessary because a spot bump otherwise changes the contract as well
as its future underlying level. The same innovations and hedge states apply to
both matched arms.

A bought option uses $-\Delta_Q$ units of stock and a sold option uses
$+\Delta_Q$, financed at the window rate and held to maturity. If $d_i$ is one
for a buy, minus one for a sell and zero otherwise,
\begin{equation}
\pi_i^{\mathrm{hedge}}=\pi_i^{\mathrm{raw}}
-d_i\Delta_{Q,i}(D_iS_{n_i}-S_{0,i})/S_{0,i}.
\end{equation}
This common initial position measures sensitivity to directional exposure for
all three contracts. The conditional path functional can affect the remaining
payoff even when the initial stock exposure is accounted for.

\begin{table}[ht]
\centering
\caption{Matched P--Q diagnostic at $g=0.05$ and $1\%S_0$ materiality. Trade
rates and P\&L are percentages. P\&L is relative to initial spot and averaged
over all 6{,}824 windows, with no trade recorded as zero.}
\label{tab:matchedpq}
\setlength{\tabcolsep}{4pt}
\small\begin{tabular}{llrrrrrr}
\toprule
Contract & $k$ & $\alpha$=1.9 trade & $\alpha$=1.9 raw & $\alpha$=1.9 hedge
& $\alpha$=2 trade & $\alpha$=2 raw & $\alpha$=2 hedge \\
\midrule
Vanilla & 0 & 39.98 & 1.293 & 0.280 & 31.92 & 0.971 & 0.181 \\
Asian & 0 & 21.23 & 0.586 & -0.155 & 15.65 & 0.449 & -0.130 \\
Lookback & 0 & 40.40 & 1.072 & 0.906 & 41.88 & -0.052 & -0.039 \\
Vanilla & 1.96 & 9.20 & 0.500 & 0.111 & 9.33 & 0.475 & 0.081 \\
Asian & 1.96 & 4.40 & 0.151 & -0.055 & 4.85 & 0.144 & -0.089 \\
Lookback & 1.96 & 7.25 & 0.410 & 0.348 & 11.78 & 0.080 & 0.063 \\
\bottomrule
\end{tabular}
\tabnote{\emph{Hedge} is one initial fixed-state Q Delta position held to
maturity. $k$ adds nominal joint Monte Carlo standard errors to the
materiality hurdle. Q means, realized paths, and hedge states are common to
both arms.}
\end{table}

\subsection{Matched terminal, average and extremum outcomes}

For $k=0$, stable-arm raw Vanilla, Asian and Lookback means are 1.293\%,
0.586\% and 1.072\% of spot. Initial hedging changes them to 0.280\%,
$-0.155\%$ and 0.906\%. With $k=1.96$, execution falls to 9.20\%, 4.40\% and
7.25\%, and hedged means become 0.111\%, $-0.055\%$ and 0.348\%. The same
generated ensembles therefore imply different execution and payoff outcomes
across the three functionals.

\begin{figure}[ht]
\centering
\includegraphics[width=\linewidth]{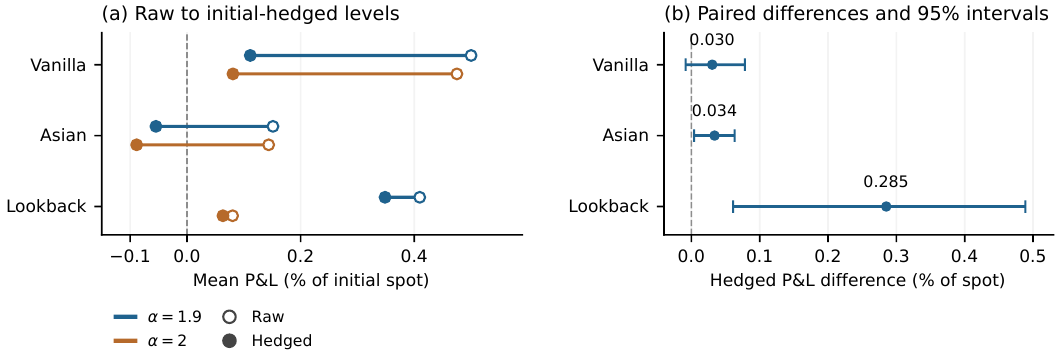}
\caption{Strict-filter economic outcomes. The first panel separates raw and
initial-hedged levels within each arm. The second reports paired hedged
differences with 60-calendar-day 95\% intervals. Values are percentages of
initial spot and average all evaluation windows, recording no trade as zero.}
\label{fig:economics}
\end{figure}

\begin{table}[t]
\centering
\caption{Paired initial-hedged differences ($\alpha=1.9$ minus $\alpha=2$) in
percent of spot. 60-day common-calendar 95\% intervals use nine nonempty
blocks.}
\label{tab:hedgeddiff}
\small\begin{tabular}{lrrr}
\toprule
Contract & $k$ & Difference & 95\% interval \\
\midrule
Vanilla & 0 & 0.099 & [-0.017, 0.215] \\
Asian & 0 & -0.025 & [-0.112, 0.062] \\
Lookback & 0 & 0.945 & [0.283, 1.660] \\
Vanilla & 1.96 & 0.030 & [-0.009, 0.078] \\
Asian & 1.96 & 0.034 & [0.004, 0.063] \\
Lookback & 1.96 & 0.285 & [0.061, 0.489] \\
\bottomrule
\end{tabular}
\end{table}

\begin{table}[t]
\centering
\caption{Initial-hedged P\&L levels under the strict quote-error filter
($k=1.96$), in percent of initial spot, averaged over all 6{,}824 windows.
Calendar-block columns give 95\% intervals, with effective block counts in
parentheses. These are level intervals; Table~\ref{tab:hedgeddiff} reports
paired differences.}
\label{tab:hedgedlevels}
\small\begin{tabular}{llrrrr}
\toprule
Arm & Contract & Mean & 60 days (9) & 120 days (5) & 252 days (3) \\
\midrule
$\alpha=1.9$ & Vanilla & 0.111 & [-0.034, 0.250] & [-0.044, 0.246] & [0.000, 0.144] \\
$\alpha=1.9$ & Asian & -0.055 & [-0.190, 0.064] & [-0.166, 0.081] & [-0.121, 0.053] \\
$\alpha=1.9$ & Lookback & 0.348 & [0.071, 0.618] & [0.020, 0.661] & [0.075, 0.444] \\
$\alpha=2$ & Vanilla & 0.081 & [-0.025, 0.179] & [-0.036, 0.183] & [0.000, 0.128] \\
$\alpha=2$ & Asian & -0.089 & [-0.221, 0.023] & [-0.185, 0.044] & [-0.159, 0.025] \\
$\alpha=2$ & Lookback & 0.063 & [-0.088, 0.223] & [-0.089, 0.263] & [-0.008, 0.187] \\
\bottomrule
\end{tabular}
\tabnote{Effective calendar-block counts are shown in the column headers. Zero
lower endpoints are shown explicitly. Option spreads are included; the stock
hedge is held to maturity without stock transaction costs.}
\end{table}

The strict-filter Vanilla hedged difference is 0.030\% of spot, with a 60-day
interval spanning zero. Asian improves by 0.034\%, with interval
$[0.004\%,0.063\%]$, while its level is negative in both arms. Lookback
improves by 0.285\%, with interval $[0.061\%,0.489\%]$; its unfiltered hedged
difference is 0.945\%, with interval $[0.283\%,1.660\%]$. These comparisons
describe distinct consequences of terminal, average and minimum-based paths,
rather than a single economic improvement shared by all contracts.

The stable-arm strict-filter Lookback level is 0.348\%, with a 60-day interval
$[0.071\%,0.618\%]$. Its level and paired-difference intervals answer
different questions. At 252 days, the stable-arm level interval remains
positive while the paired interval includes zero. Strict-filter Lookback
difference intervals also span zero at 90 and 150 days.

\begin{table}[p]
\singlespacing
\centering
\caption{Strict-filter paired initial-hedged differences by calendar
resolution}
\label{tab:hedgedblocks}
\small\begin{tabular}{lrr}
\toprule
Contract & Block days & Difference CI (\% of spot) \\
\midrule
Vanilla & 20 & [-0.005, 0.069] \\
Vanilla & 60 & [-0.009, 0.078] \\
Vanilla & 90 & [-0.014, 0.069] \\
Vanilla & 120 & [-0.014, 0.077] \\
Vanilla & 150 & [-0.011, 0.085] \\
Vanilla & 180 & [-0.006, 0.084] \\
Vanilla & 252 & [0.000, 0.055] \\
Asian & 20 & [0.005, 0.063] \\
Asian & 60 & [0.004, 0.063] \\
Asian & 90 & [0.009, 0.052] \\
Asian & 120 & [0.010, 0.051] \\
Asian & 150 & [-0.001, 0.049] \\
Asian & 180 & [0.000, 0.046] \\
Asian & 252 & [0.000, 0.038] \\
Lookback & 20 & [0.115, 0.444] \\
Lookback & 60 & [0.061, 0.489] \\
Lookback & 90 & [-0.044, 0.569] \\
Lookback & 120 & [0.073, 0.561] \\
Lookback & 150 & [-0.046, 0.541] \\
Lookback & 180 & [0.035, 0.398] \\
Lookback & 252 & [-0.112, 0.452] \\
\bottomrule
\end{tabular}
\tabnote{Differences are $\alpha=1.9$ minus $\alpha=2$ at $k=1.96$. Intervals
quantify differences between the arms in their mean initial-hedged P\&L. The
same calendar resampling and window weighting apply at every resolution.}
\end{table}

\begin{table}[p]
\centering
\caption{Corrected-input original-model economic reference}
\label{tab:drefecon}
\small\begin{tabular}{lrrrr}
\toprule
Contract & $k$ & Execution (\%) & Raw (\%) & Hedged (\%) \\
\midrule
Vanilla & 0 & 31.70 & 0.831 & 0.392 \\
Vanilla & 1.96 & 4.16 & 0.257 & 0.088 \\
Asian & 0 & 12.82 & 0.311 & -0.024 \\
Asian & 1.96 & 1.54 & 0.039 & -0.035 \\
Lookback & 0 & 37.47 & 0.183 & 0.192 \\
Lookback & 1.96 & 7.71 & 0.073 & 0.074 \\
\bottomrule
\end{tabular}
\tabnote{All means include 6{,}824 windows, no trade as zero, and normalize by
initial spot. $\Dref$ uses the original trained weights with corrected
histories and fixed preprocessing. The initial hedge fixes observed contract
state and includes financing.}
\end{table}

\subsection{Original-model economics, quote concentration and drift}

Corrected-input $\Dref$ raw Vanilla, Asian and Lookback means are 0.831\%,
0.311\% and 0.183\%; their fixed-state initial-hedged means are 0.392\%,
$-0.024\%$ and 0.192\%. Original-model Lookback raw and hedged level intervals
span zero across 20--150-day blocks. Its buy contribution is $+0.559\%$ and its
sell contribution $-0.376\%$, with sells accounting for 75.40\% of executed
trades.

Nested subsets of the original model's recorded ensemble use the first 16, 32
and all 40 paths. At $k=1.96$, raw Vanilla means are 0.165\%, 0.220\% and
0.257\%; Lookback means are 0.016\%, 0.069\% and 0.073\%. The nested
diagnostic quantifies the effect of ensemble size on quote concentration and
threshold crossings.

Terminal-drift alignment uses the same original-model paths and multiplies a
path by $\exp(aj/n)$, where
\begin{equation}
a=\log\left(\frac{e^{rn/252}}{\overline{S_n/S_0}}\right).
\end{equation}
This sets the ensemble discounted terminal mean to one, then recomputes P
quotes and decisions. The resulting raw means are $-0.954\%$, $-0.315\%$ and
$-1.660\%$ for Vanilla, Asian and Lookback. In contrast, the initial hedge
retains the original quotes and decisions. The two diagnostics therefore
isolate different consequences of physical direction: one changes the price
path supplied to the payoff, and the other changes the realized accounting for
the already chosen trade.

\subsection{Structured-product conventions}

Accumulator and snowball scenarios extend the path-functional evaluation to
repeated fixings, leverage and early termination. Their definitions are held
fixed when the existing original and matched ensembles are rescored. The
section evaluates undiscounted, normalized-notional cashflow distributions and
event frequencies.

For the accumulator, let $R_j=S_j/S_0$ and let $\tau_A$ be the first future
observation with $R_j\ge 1.05$, or $n+1$ if none occurs. The scenario cashflow
is
\begin{equation}
H_A^{\mathrm{acc}}=\frac{1}{n}\sum_{j=1}^{n}\mathbf{1}\{j<\tau_A\}
\,(1+\mathbf{1}\{R_j<0.85\})\,(R_j-0.85).
\end{equation}
Observations are daily, quantity doubles below the 0.85 strike, and the first
KO fixing itself does not accrue. The denominator counts all scheduled future
observations.

For the snowball, KO is observed every fifth trading observation and at
maturity. Let $\tau_B$ be the first such observation with $R_j\ge 1.05$; let
$I=\mathbf{1}\{\min_{1\le j\le n}R_j\le 0.80\}$. The implemented cashflow is
\begin{equation}
H_B^{\mathrm{snow}}=
\begin{cases}
0.15\tau_B/252, & \tau_B\le n,\\[4pt]
0.15n/252, & \text{no KO and } I=0,\\[4pt]
\min(R_n-1,0), & \text{no KO and } I=1.
\end{cases}
\end{equation}
KO takes precedence. A path with a barrier crossing, no KO and terminal
recovery to initial spot receives zero under this scenario convention. The
reported barrier statistic examines the complete future path, including
observations after a preceding KO; it is a full-horizon barrier-crossing
diagnostic, distinct from contractual KI before termination. These
definitions specify the archived implementation and are applied unchanged to
all distributions.

\subsection{Structured cashflows, events and matched consequences}

Table~\ref{tab:structured} applies the fixed contracts to the original
recorded-input model, corrected $\Dref$ and both fresh matched arms over the
same 6{,}824 windows. Each P ensemble contains 40 paths; all comparisons use
the same 256 Q paths per window and the same realized cashflows.

\begin{table}[t]
\singlespacing
\centering
\caption{Structured-product scenarios from the archived and corrected path
ensembles}
\label{tab:structured}
\small\begin{tabular}{llrrrr}
\toprule
Scenario & Distribution & Mean (\%) & ES$_5$ (\%) & KO (\%) & Loss (\%) \\
\midrule
Accumulator & Realized & 9.334 & 0.604 & 52.71 & 0.015 \\
Accumulator & Q & 9.805 & -0.891 & 43.50 & 1.018 \\
Accumulator & Original archive & 11.012 & 0.906 & 45.82 & 0.006 \\
Accumulator & $\Dref$ & 11.091 & 0.919 & 44.26 & 0.009 \\
Accumulator & $\alpha=1.9$ & 10.807 & 0.906 & 49.29 & 0.005 \\
Accumulator & $\alpha=2$ & 12.272 & 2.066 & 42.96 & 0.000 \\
Snowball & Realized & 1.936 & -2.494 & 49.68 & 1.495 \\
Snowball & Q & 1.862 & -9.965 & 39.09 & 2.461 \\
Snowball & Original archive & 2.450 & 0.370 & 42.40 & 0.036 \\
Snowball & $\Dref$ & 2.497 & 0.343 & 40.82 & 0.044 \\
Snowball & $\alpha=1.9$ & 2.362 & 0.422 & 46.00 & 0.023 \\
Snowball & $\alpha=2$ & 2.532 & 0.786 & 40.34 & 0.000 \\
\bottomrule
\end{tabular}
\tabnote{Cashflows and ES$_5$ use undiscounted normalized-notional units.
ES$_5$ is the pooled path-payoff mean at or below the empirical 5\% quantile.
Each P ensemble has 40 paths per window, Q has 256; realized paths provide one
observation per window. Original archive is distinguished from corrected
$\Dref$.}
\end{table}

The matched accumulator mean is 10.807\% of notional for $\alpha=1.9$ and
12.272\% for $\alpha=2$, compared with 9.334\% realized and 9.805\% under Q.
The stable arm also raises KO frequency from 42.96\% to 49.29\%, closer to the
realized 52.71\%. For snowball, matched means are 2.362\% and 2.532\%,
compared with 1.936\% realized. Its KO frequency rises from 40.34\% to
46.00\%, compared with 49.68\% realized. More frequent early termination
changes both the cashflow horizon and coupon accrual under these definitions.

\begin{table}[ht]
\centering
\caption{Snowball full-horizon barrier and coupon frequencies}
\label{tab:snowballevents}
\small\begin{tabular}{lrr}
\toprule
Distribution & Barrier crossing (\%) & Positive coupon (\%) \\
\midrule
Realized & 1.964 & 98.491 \\
Q & 3.294 & 97.518 \\
Original archive & 0.037 & 99.964 \\
$\Dref$ & 0.045 & 99.956 \\
$\alpha=1.9$ & 0.024 & 99.977 \\
$\alpha=2$ & 0.000 & 100.000 \\
\bottomrule
\end{tabular}
\tabnote{The barrier statistic checks whether $S_j\le 0.80S_0$ anywhere in the
complete recorded future path, including observations after a preceding KO.
It therefore measures full-horizon barrier crossing, not contractual KI before
termination. KO takes precedence in the payoff. Coupon denotes a strictly
positive scenario cashflow.}
\end{table}

The full-horizon snowball barrier frequencies are 0.024\% for the stable arm,
zero in the Gaussian ensemble, 1.964\% realized and 3.294\% under Q. Both
neural ensembles underrepresent full-horizon barrier crossings and downside
loss paths relative to the realized panel. These event and lower-tail
statistics add multi-observation path properties to the first-return tail
diagnostics.

\begin{table}[ht]
\centering
\caption{Matched structured-scenario differences}
\label{tab:structureddiff}
\small\begin{tabular}{llrrr}
\toprule
Scenario & Statistic & Difference & 60-day CI & 252-day CI \\
\midrule
Accumulator & Mean cashflow & -1.465 & [-1.511, -1.419] & [-1.491, -1.333] \\
Accumulator & KO frequency & 6.336 & [5.622, 7.492] & [5.503, 11.033] \\
Accumulator & Absolute payoff error & -0.435 & [-0.692, -0.105] & [-0.478, 1.033] \\
Snowball & Mean cashflow & -0.170 & [-0.193, -0.142] & [-0.191, -0.050] \\
Snowball & KO frequency & 5.668 & [4.966, 6.780] & [4.847, 10.435] \\
Snowball & Absolute payoff error & -0.001 & [-0.051, 0.053] & [-0.009, 0.044] \\
\bottomrule
\end{tabular}
\tabnote{Differences are $\alpha=1.9$ minus $\alpha=2$, in percentage points
of notional for cashflows/errors and percentage points for KO frequency.
Absolute payoff error is $|\widehat{\mathbb{E}}[H_i]-H_i^{\mathrm{realized}}|$
per window. Negative absolute-payoff-error differences favor $\alpha=1.9$.
Intervals use common calendar blocks.}
\end{table}

For a window-level cashflow forecast $\widehat\mu_i$, define the diagnostic
absolute payoff error as $|\widehat\mu_i-H_i^{\mathrm{realized}}|$. The
stable-minus-Gaussian accumulator error difference is $-0.435$ percentage
points of notional, with 60-day interval $[-0.692,-0.105]$. The corresponding
snowball error difference is $-0.001$ percentage points, with interval
$[-0.051,0.053]$. The accumulator error-difference interval spans zero at 252
days, making its uncertainty sensitive to calendar resolution. These scenario
diagnostics complement the smooth-option P\&L results in their own economic
units.

\section{Discussion and Conclusion}\label{sec:conclusion}

The study develops a history-aware financial path generator and evaluates the
same conditional distributions through proper forecast scores and nonlinear
cashflows. Coordinate-wise stable mixing, a conditional U-Net and financial
regularization adapt DLPM to equity-index paths. The common evaluation links
historical forecasts to nonlinear cashflows and realized outcomes under shared
Q quotes.

With initialization, training information, loss and budget fixed, the
$\alpha=1.9$ model reduces terminal CRPS by 5.05\% and path energy by 4.03\%
relative to $\alpha=2$. Coverage and interval score improve together, and the
bounded outputs retain more frequent first-return extremes. The treatment
includes stable mixing and its associated scale-preserving coefficients under
the common numerical implementation. Historical terminal pooling and complete
path resampling remain competitive references; neighborhood-size sensitivity
places the original model's local-control advantage in that wider context.

Payoff functionals reveal different consequences of the generated paths. The
strict-filter initial-hedged Lookback difference is 0.285\% of spot, with a
60-day interval $[0.061\%,0.489\%]$. Vanilla's difference interval spans zero;
Asian improves relative to the Gaussian arm while its levels stay negative.
Calendar resolution affects the Lookback difference intervals. Accumulator and
snowball scenarios extend the comparison to termination and barrier events.
The stable arm improves the accumulator absolute payoff-error diagnostic,
while the snowball error difference interval spans zero and its downside event
frequencies remain below the realized panel. These distinctions show why
terminal scores, extrema, averages and structured cashflows should be
evaluated together.

The conclusions are conditional on one matched training seed, a
retrospectively extended historical panel and fixed bounded sampling.
Common-calendar intervals describe variation within that panel, with
overlapping future targets across origin-date blocks and few groups at long
resolutions. Future work can extend the matched design across seeds and market
periods, examine calibration and history-conditioning alternatives, and
connect the payoff evaluation to observed option quotes, carry data and
dynamically rebalanced hedges. Structured-product extensions can additionally
target live-contract barrier probabilities and downside cashflow
distributions.

\paragraph{Acknowledgment.}
OpenAI Codex assisted with manuscript drafting and editing, preparation of
tables and figures from recorded results, and experimental scripting and
verification. The authors are responsible for the scientific claims and final
manuscript.

\clearpage
\appendix

\section{Network Architecture}\label{app:architecture}

\paragraph{Backbone.} We use a one-dimensional conditional U-Net
\citep{ronneberger2015unet} with base width 64, channel multipliers
$(1, 2, 4, 8)$, dropout 0.1. The prediction output is bounded by
$2\tanh(\cdot)$. The encoder consists of residual convolutional blocks and
downsampling layers that halve sequence length while increasing feature
dimension; skip connections preserve encoder features for the decoder, which
restores resolution by upsampling and concatenation. Linear attention operates
at each resolution and full self-attention at the bottleneck. Each residual
block is
\begin{equation}
\mathrm{ResBlock}(x) = x +
\mathrm{Conv}\bigl(\mathrm{SiLU}(\mathrm{Norm}(\mathrm{Conv}(x)))\bigr),
\end{equation}
with conditioning injected via feature-wise scale--shift modulation.

\paragraph{Time embedding.} Diffusion steps enter through sinusoidal
embeddings \citep{vaswani2017attention},
\begin{equation}
\mathrm{TimeEmb}(t) = \bigl[\sin(\omega_1 t), \cos(\omega_1 t), \ldots,
\sin(\omega_{d/2}\, t), \cos(\omega_{d/2}\, t)\bigr],
\qquad \omega_i = 10000^{-2i/d}.
\end{equation}

\paragraph{Condition embedding.} The two categorical identifiers pass through
learned embedding tables (country: 64 dimensions; underlying index: 128
dimensions). Temporal encoders separately process the 60-day return sequence,
252-day return sequence and 60-day volatility path, with 32-dimensional
history representations. The numerical stream contains the contract variables
and history-aware features: 60/252-day returns, trend, drawdown, realized and
downside volatility, volatility paths, reversal and validity indicators. It is
projected with LayerNorm and SiLU; the streams are fused by learned weighted
concatenation and a two-layer MLP into a 128-dimensional condition vector,
\begin{equation}
\mathrm{CondEmb}(c) = \mathrm{MLP}\bigl(
[\,w_n\,\phi_n(c_{\mathrm{num}});\;
  w_c\,E_{\mathrm{cty}}(c_{\mathrm{country}});\;
  w_x\,E_{\mathrm{idx}}(c_{\mathrm{index}})\,]\bigr) \in \mathbb{R}^{128},
\end{equation}
which modulates every residual block together with the time embedding. The
index embedding supplies underlying identity within the shared cross-index
model.

\section{Training and Sampling Hyperparameters}\label{app:hyper}

\begin{table}[ht]
\singlespacing
\centering
\caption{Shared fresh matched-training and generation recipe}
\label{tab:hyper}
\small\begin{tabular}{>{\raggedright\arraybackslash}p{0.34\linewidth}>{\raggedright\arraybackslash}p{0.58\linewidth}}
\toprule
Setting & Value \\
\midrule
Stability indices & 1.9 and 2, canonical $\mathcal{S}\alpha\mathcal{S}(1)$ normalization \\
Training windows / batch / updates & 59{,}240 / 32 / 16{,}000 \\
Training seed / shuffle & 20260730; independent CPU epoch seed $20261009+e$ \\
Checkpoint & EMA-12{,}000, fixed before matched runs \\
Optimizer & Adam, betas $(0.9,0.99)$ \\
Learning-rate schedule & 2{,}400-step linear warm-up, 1\% floor; $10^{-5}$ peak; cosine decay to $10^{-7}$ \\
EMA & 0.995 decay; update every ten steps \\
Forward diffusion steps & 1{,}000, cosine-derived scale-preserving schedule \\
Target / return scale & 252-coordinate masked tensor / 0.09 \\
Denoising objective & Masked Smooth L1; valid-coordinate denominator; regime-weighted batch mean \\
Auxiliary objective & Fixed coefficients; scale 0.07; 1{,}000-update warm-up \\
Mixture cap & 100 for training and reverse mixture chains \\
Prediction / generation bounds & Noise prediction $[-2,2]$; initial noise $[-2,2]$; denoised/final tokens $[-1.35,1.35]$ \\
Intermediate trust bound & $8\bar\gamma_t+2\bar\sigma_t$ \\
Auxiliary feature clip / range penalty & 2.5 / 1.25 \\
Sampling & 50 skip steps, $\eta=1$; 40 P paths per window \\
P streams & 32 paths seed 20260730; 8 paths seed 20260803 \\
Q ensemble / carry & 256 paths; zero carry under common GARCH benchmark \\
Calendar intervals & 2{,}000 percentile replicates; blocks 20--252 calendar days \\
\bottomrule
\end{tabular}
\end{table}

\section{Reproducibility and Data Availability}\label{app:repro}

The release supplies editable LaTeX, every included table and figure, the
machine-readable path/economic summaries, the matched training protocol,
independent scoring verification and structured-scenario window statistics.
Complete source and artifact hashes are recorded in the delivery manifest.
The accompanying verification separates original recorded-input outputs,
corrected $\Dref$ and the fresh matched arms. It also records the stored Q
path files used for the structured evaluation.

The corrected-input and matched experiments share the historical targets,
rates and 6{,}824-window calendar. The supplied protocols document training,
checkpoint selection and evaluation settings.

Training seed 20260730 is common to the fresh arms. The archival
neural-control families report available runs with seeds 20260730--20260732.
Sampling streams and their mixture-cache schedule are specified in the
protocol. Raw market series remain subject to provider redistribution terms;
source tickers, transforms, chronological boundaries and checksums document
reconstruction.

\clearpage
\singlespacing

\end{document}